\documentclass[12pt, english]{article}
\usepackage[utf8]{inputenc}
\usepackage{float}
\usepackage{bbm}
\usepackage{advdate}
\usepackage{amsmath}
\usepackage[ruled,vlined]{algorithm2e}
\usepackage{amssymb}
\usepackage{url}
\usepackage{bm}
\usepackage[colorlinks=true, allcolors=blue]{hyperref}
\usepackage[round]{natbib}
\usepackage{booktabs}
\usepackage{caption}
\usepackage{subcaption}
\usepackage{graphicx}
\usepackage{geometry}
\newtheorem{assumption}{Assumption}[section]
\usepackage{pdflscape}
\usepackage{pgfplots}
\pgfplotsset{compat=1.7}
\usepackage[flushleft]{threeparttable}
\usepackage[flushleft]{threeparttablex}
\usepackage{array}
\usepackage{rotating}
\usepackage{longtable}

\usepackage{enumitem}
\usepackage[toc,page]{appendix}
\newenvironment{fignote}{\begin{quote}\footnotesize}{\end{quote}}
\usepackage{xcolor}
\usepackage{multirow}
\usepackage{authblk}

\usepackage{tikz}
\usepackage{tabularx}
\usetikzlibrary{decorations.pathreplacing}
\usepackage{adjustbox}

\usepackage{babel}
\title{\textbf{A Way to Synthetic Triple Difference\thanks{This work was supported by Peking University School of Economics Research Seed Grant, and the Fundamental Research Funds for the Central Universities, Peking University (7100604568). I am grateful for helpful comments from Wenli Xu and an associate editor from Journal of Econometrics. All errors are nevertheless my own.}}}
\author{Castiel Chen Zhuang\thanks{zogcee@gmail.com, Peking University School of Economics.}}
\date{First version: September 17, 2024 \\This version: September 22, 2024}

\begin{document}

\maketitle
\vspace{-5mm}
\begin{abstract}
    This paper discusses a practical approach that combines synthetic control with triple difference to address violations of the parallel trends assumption. By transforming triple difference into a DID structure, we can apply synthetic control to a triple-difference framework, enabling more robust estimates when parallel trends are violated across multiple dimensions. The proposed procedure is applied to a real-world dataset to illustrate when and how we should apply this practice, while cautions are presented afterwards. This method contributes to improving causal inference in policy evaluations and offers a valuable tool for researchers dealing with heterogeneous treatment effects across subgroups.
    
    \vspace{2mm}
    
    \vspace{2mm}
\end{abstract}

\newpage
\section{Introduction}
In recent empirical research, the difference-in-differences (DID) and triple-difference (i.e., difference-in-difference-in-differences, or DDD hereinafter) estimators have been widely adopted for estimating causal effects due to their intuitive design and ability to control for unobserved time-invariant heterogeneity across units. However, one key challenge that researchers face is ensuring that the parallel trends assumption holds. Several strategies have been developed to facilitate or approximate parallel trends in DID frameworks, including combining propensity score matching (PSM), entropy balancing, and synthetic control with DID \citep{hainmueller2012, arkhangelsky2021}. These methods improve the comparability of treatment and control groups by reweighting or constructing synthetic controls that more closely mimic the characteristics of the treated group. However, despite the attention paid to improving parallel trends in DID, there has been little work devoted to addressing this issue within the DDD framework.

In this paper, I aim to initiate a conversation with the literature by proposing an approach that combines synthetic control (SC) with DDD. The SC method, originally designed to improve the counterfactual for DID settings \citep{abadie2010synthetic}, offers a powerful tool for constructing better control groups for DDD as well. However, its application to DDD settings, where treatment effects may differ across subgroups, has been underexplored. By extending the SC approach to a transformed version of DDD, I show how it can improve the validity of estimates when parallel trends are violated across multiple dimensions (including treatment status, time, and subgroup characteristics) due to the exponential growth of the outcome value.\footnote{In such situation, a synthetic DDD estimator may be needed when taking the logarithm is not a good idea, e.g., if the outcome takes the value of zero for some units.} A synthetic DDD estimator opens an avenue for causal inference when synthetic DID still performs poorly.

The contribution of this paper is two-fold. First, I contribute to the discussion of the SC methodology in DDD settings, showing how transforming DDD into DID can help facilitate parallel trends in more complex empirical applications. Second, I provide an empirical example to provide a practical guide for this approach in a real-world policy evaluation; moreover, in this empirical exercise, I compare the performance of this synthetic DDD method to standard DDD and DID estimators, illustrating the effectiveness and highlighting the potential benefits of this combined approach in improving the robustness of causal inference.

The remainder of this paper is structured as follows. In Section 2, I present the theoretical foundations of the transformed DDD model. Section 3 explains how SC can be adapted for use in a DDD context based on an empirical example, including a brief explanation of the dataset and the policy under study; the results of the empirical analysis are discussed, and the performances of standard DDD, DID, and the proposed synthetic DDD estimators are compared. In Section 4, I lay out additional notes and cautions for readers who intend to apply this procedure. Finally, Section 5 concludes with a discussion of the broader implications for empirical research and policy evaluation.

\section{Simplifying an Estimation by Demeaning}

\subsection{A simple illustration}
Let us start with a simple DID model without covariates. I would like to demonstrate how a DID estimator can be obtained by transforming the outcome variable through subtracting the control group's mean outcome at each time period. This approach allows us to estimate treatment effects using simplified regressions without covariates or fixed effects.

Consider the standard DID model with fixed effects:
\begin{equation}
    Y_{it} = \alpha_0 + \alpha_1 (\text{Treat}_i \times \text{Post}_t) + \mu_i + \lambda_t + \epsilon_{it},
\end{equation}
where $Y_{it}$ is the outcome variable for unit $i$ at time $t$, $\text{Treat}_i$ is the indicator variable equal to 1 if unit $i$ is in the treatment group and 0 otherwise, $\text{Post}_t$ is the indicator variable equal to 1 if time $t$ is in the post-treatment period and 0 otherwise, $\mu_i$ refers to a unit (individual) fixed effect, $\lambda_t$ refers to a time fixed effect, $\alpha_0$ is the common intercept term, $\alpha_1$ is the average treatment effect on the treated (ATT), which is the parameter of interest, and $\epsilon_{it}$ is the error term.

Suppose that Assumptions \ref{ass:parallel1} and \ref{ass:exogeneity1} are satisfied.

\begin{assumption} \label{ass:parallel1}
    \textbf{Parallel Trends Assumption}: In the absence of treatment, the expected change in outcomes over time is the same for both treatment and control groups:
    \begin{align*}
        &\mathbb{E}[Y_{0,it} \mid \text{Treat}_i = 1, \text{Post}_t = 1] - \mathbb{E}[Y_{0,it} \mid \text{Treat}_i = 1, \text{Post}_t = 0] \\ = &\mathbb{E}[Y_{0,it} \mid \text{Treat}_i = 0, \text{Post}_t = 1] - \mathbb{E}[Y_{0,it} \mid \text{Treat}_i = 0, \text{Post}_t = 0].
    \end{align*}
    where $Y_0$ follows the notation of \citet{angrist2009}'s potential outcome framework.
\end{assumption}
\begin{assumption} \label{ass:exogeneity1}
    \textbf{Exogeneity}: $\mathbb{E}[\varepsilon_{it} \mid \text{Treat}_i, \text{Post}_t, \mu_i, \lambda_t] = 0$.
\end{assumption}

Then, the ATT can be written as
\begin{align}
\begin{split}
    \alpha_1 = &\left( \mathbb{E}[Y_{it} \mid \text{Treat}_i = 1, \text{Post}_t = 1] - \mathbb{E}[Y_{it} \mid \text{Treat}_i = 1, \text{Post}_t = 0] \right) - \\
    &\left( \mathbb{E}[Y_{it} \mid \text{Treat}_i = 0, \text{Post}_t = 1] - \mathbb{E}[Y_{it} \mid \text{Treat}_i = 0, \text{Post}_t = 0] \right). \label{eq:att}
\end{split}
\end{align}

As shown in Appendix \ref{proof:att}, this is equivalent to $\beta_1$ in the following model for individuals in the treatment group ($\text{Treat}_i = 1$):
\begin{equation}
    Z_{it} = \beta_0 + \beta_1 \text{Post}_t + u_{it} \text{, if } \text{Treat}_i = 1,
\end{equation}
where we define the transformed outcome variable as
\begin{equation}
    Z_{it} = Y_{it} - \mathbb{E}[Y_{it} \mid \text{Treat}_i = 0, t = t] \label{eq:z}
\end{equation}
and $\mathbb{E}[Y_{it} \mid \text{Treat}_i = 0, t = t] = \bar{Y}_{\text{Control}, t}$ as the expected outcome for the control group at time $t$. In practice, we can estimate $\bar{Y}_{\text{Control}, t}$ using sample averages:
\begin{equation}
    \hat{\bar{Y}}_{\text{Control}, t} = \frac{1}{\sum_{i} 1\{\text{Treat}_i = 0\}} \sum_{i \in \{i | \text{Treat}_i = 0\}} Y_{it}.
\end{equation}

Thus, the sample counterpart of $Z_{it}$ is:
\begin{equation}
    \hat{Z}_{it} = Y_{it} - \hat{\bar{Y}}_{\text{Control}, t}.
\end{equation}

We estimate the following regression for the treated group ($\text{Treat}_i = 1$):
\begin{equation}
    \hat{Z}_{it} = \hat{\beta}_0 + \hat{\beta}_1 \text{Post}_t + \hat{u}_{it} \text{, if } \text{Treat}_i = 1,
\end{equation}
where $\hat{\beta}_1$ is the estimator for the treatment effect $\alpha_1$ and $\hat{u}_{it}$ is the residual term.

Under Assumptions \ref{ass:parallel1} and \ref{ass:exogeneity1}, $\hat{\beta}_1$ is a consistent estimator of $\alpha_1$:
\begin{equation}
    \hat{\beta}_1 = \left( \hat{\bar{Y}}_{\text{Treat}, \text{Post}} - \hat{\bar{Y}}_{\text{Control}, \text{Post}} \right) - \left( \hat{\bar{Y}}_{\text{Treat}, \text{Pre}} - \hat{\bar{Y}}_{\text{Control}, \text{Pre}} \right) \xrightarrow{p} \alpha_1,
\end{equation}
where $\hat{\bar{Y}}_{\text{Group}, \text{Time}}$ denotes sample averages for Group (either treatment or control groups) and Time (either pre-treatment or post-treatment years), and ``$\xrightarrow{p}$'' denotes convergence in probability.

Note that, this method does not involve partialling out variables using regression residuals as in the Frisch-Waugh-Lovell (FWL) theorem. Instead, it simply adjusts the outcome variable by subtracting sample estimates of control group means at each time period. Despite not following the requirements of the FWL theorem, this method aims to control for confounding factors (e.g., time effects, covariates) by adjusting the outcome variable. The validity of the estimator relies on the assumptions stated earlier (e.g., parallel trends) and correct specification of the model.

\subsection{Converting DDD to DID}

Let us first consider a standard triple-difference model with fixed effects but without covariates as follows:\footnote{Note that, this is a fully saturated model as in \citet{olden2022}, since $Treat_i \times G_i$, $Treat_i$, and $G_i$ are absorbed by $\mu_i$, while $Post_t$ is absorbed by $\lambda_i$. One may also consider heterogeneous treatment effect by replacing $\delta$ by $\delta_i$}
\begin{equation}
    Y_{it} = \alpha + \delta D_{it} \times G_i + \gamma_1 D_{it} + \gamma_2 G_i \times Post_t + \mu_i + \lambda_t + \varepsilon_{it},
\end{equation}
where $D_{it} = \text{Treat}_i \times \text{Post}_t$ is the DID interaction term, $G_i$ is the indicator variable equal to 1 if individual $i$ is in the subgroup of interest (such as a policy target) and 0 otherwise, $\delta$ is the DDD effect, and $\gamma_1$ and $\gamma_2$ are coefficients capturing interaction effects.

For individuals in the subgroup of interest ($G_i = 1$), we can define the transformed outcome variable in $\text{Treat}_i = j$ ($j = 0,1$) as
\begin{equation}
    W_{itj} = Y_{it} - \mathbb{E}[Y_{it} \mid G_i = 0, \text{Treat}_i = j, t = t],
\end{equation}
where $\mathbb{E}[Y_{it} \mid G_i = 0, \text{Treat}_i = j, t = t] = \bar{Y}_{\text{Non-target}, j, t}$ is the expected outcome for the subgroup of interest in $\text{Treat}_i = j$ ($j = 0,1$) at time $t$. In practice, we can estimate $\bar{Y}_{\text{Non-target}, j, t}$ using sample averages:
\begin{equation}
    \hat{\bar{Y}}_{\text{Non-target}, j, t} = \frac{1}{\sum_{i} 1\{G_i = 0, \text{Treat}_i = j\}} \sum_{i \in \{i | G_i = 0, \text{Treat}_i = j\}} Y_{it} \text{ for $j=0,1$}.
\end{equation}

Thus, the sample counterpart of $W_{itj}$ is:
\begin{equation}
    \hat{W}_{itj} = Y_{it} - \hat{\bar{Y}}_{\text{Non-target}, j, t}. \label{eq:w}
\end{equation}

We estimate the following DID regression for individuals with $G_i = 1$:
\begin{equation}
    \hat{W}_{itj} = \hat{\theta}_0 + \hat{\theta}_1 D_{it} + \hat{\mu}_i + \hat{\lambda}_t + \hat{e}_{it} \text{, if } G_i = 1,
\end{equation}
where $\hat{\theta}_1$ is the estimator for $\delta$. As shown in Appendix \ref{proof:ddd}, under Assumptions \ref{ass:parallel2} and \ref{ass:exogeneity2}, $\hat{\theta}_1$ is a consistent estimator of $\delta$.

\begin{assumption} \label{ass:parallel2}
    \textbf{Extended Parallel Trends Assumption}: \citet{olden2022} point out that we only need one parallel trends assumption rather than two. This assumption can be formally written as:
    \begin{align*}
        &\left[ (\mathbb{E}[Y_{0,it} \mid \text{T}_i = 1, \text{P}_t = 1, \text{G}_i = 1] - \mathbb{E}[Y_{0,it} \mid \text{T}_i = 1, \text{P}_t = 0, \text{G}_i = 1]) \right. \\
        &- \left. (\mathbb{E}[Y_{0,it} \mid \text{T}_i = 0, \text{P}_t = 1, \text{G}_i = 1] - \mathbb{E}[Y_{0,it} \mid \text{T}_i = 0, \text{P}_t = 0, \text{G}_i = 1]) \right] \\
        = & \left[ (\mathbb{E}[Y_{0,it} \mid \text{T}_i = 1, \text{P}_t = 1, \text{G}_i = 0] - \mathbb{E}[Y_{0,it} \mid \text{T}_i = 1, \text{P}_t = 0, \text{G}_i = 0]) \right. \\
        &- \left. (\mathbb{E}[Y_{0,it} \mid \text{T}_i = 0, \text{P}_t = 1, \text{G}_i = 0] - \mathbb{E}[Y_{0,it} \mid \text{T}_i = 0, \text{P}_t = 0, \text{G}_i = 0]) \right].
    \end{align*}
\end{assumption}
\begin{assumption} \label{ass:exogeneity2}
    \textbf{Extended Exogeneity}: The error term $\varepsilon_{it}$ must satisfy $\mathbb{E}[\varepsilon_{it} \mid \text{Treat}_i, \text{Post}_t, \text{Group}_i, \mu_i, \lambda_t] = 0$.
\end{assumption}

With the above derivations, it is not hard to consider the case when there are covariates $X_{it}$. For individuals with $G_i = 1$, we adjust the outcome variable as
\begin{equation}
    W_{itjx} = Y_{it} - \mathbb{E}[Y_{it} \mid G_i = 0, \text{Treat}_i=j, t = t, X_{it} = x]. \label{eq:fit}
\end{equation}

In practice, since $X_{it}$ usually involves continuous variables, we may not be able to estimate $\mathbb{E}[Y_{it} \mid G_i = 0, \text{Treat}_i = j, t = t, X_{it} = x]$ using unconditional means in subgroups but regression models fitted on the non-target subgroup ($G_i = 0$). The linear case is given by
\begin{equation}
    \hat{Y}_{it} = X^{\prime}_{it} \hat{\beta}_{\text{Non-target},j,t} \text{ for $\text{Treat}_i = j = 0,1$ and $t$}. 
\end{equation}

Thus, the sample counterpart of the adjusted outcome variable is:
\begin{equation}
    \hat{W}_{itjx} = Y_{it} - x^{\prime} \hat{\beta}_{\text{Non-target},j,t}, \label{eq:w2}
\end{equation}
where $x$ is a specific value of $X_{it}$. We then estimate the following regression for individuals with $G_i = 1$:
\begin{equation}
    \hat{W}_{itjx} = \hat{\theta}_0 + \hat{\theta}_1 D_{it} + \hat{\mu}_i + \hat{\lambda}_t + \hat{e}_{it} \text{, if } G_i = 1,
\end{equation}
with $\hat{\theta}_1$ being a consistent estimator for the DDD effect with covariates, under the assumption that the relationship between $X_{it}$ and $Y_{it}$ is correctly specified and stable across groups.

After transforming the DDD estimator to a DID estimator by demeaning, we are ready to discuss combining DDD with SC in practice, when the extended parallel trends assumption with covariates is violated.

\section{Opening an Avenue for Synthetic DDD}

As suggested by \citet{olden2022}, all procedures for DID can be applied to a transformed DDD. This opens an avenue for us to combine SC with triple difference based on \citet{arkhangelsky2021}. When should we apply this? How exactly can we do so in empirical research and how does it perform? In the following subsections, I answer these questions based on an empirical example.

\subsection{When?}

Suppose that we are interested in studying the impact of the ``4+7 Cities'' Centralized Drug Procurement pilot program initiated at the end of 2018 (see Appendix \ref{appendix:policy} for more details) on medical and drug-related innovation in China, and we have a dataset that consists of the number of patents by the first four digits of International Patent Classification symbol (i.e., four-digit IPC code), patent type, and inventor type in each city every four year from 2001 (i.e., 2001, 2005, 2009, 2013, 2017, and 2021). A natural idea is to estimate the following equation:
\begin{equation}
    \text{NumPatents}_{it} = \alpha_0 + \alpha_1 (\text{Treat}_i \times \text{Post}_t) + X^{\prime}_{it} \alpha_2 + \mu_i + \lambda_t + \epsilon_{it},
\end{equation}
where $i$ denotes a ``city, four-digit IPC code, patent type, and inventor type'' pair (i.e., a track), $t$ denotes a year, and $\text{NumPatents}_{it}$ is the number of patent applications in a specific track and year. $\text{Treat}_i$ takes the value of 1 if a track is in one of the ``4+7'' pilot cities, while $\text{Post}_t$ takes the value of 1 after year 2018. Since the policy targets the medical and drug-related fields, we may limit our analysis to patents whose IPC symbol starts with ``A61''. Based on the data I compiled from China National Intellectual Property Administration (see more details from Appendix \ref{appendix:data}), there are 2,058 observations for A61 tracks and 11,478 observations for non-A61 tracks. I only include observations with a positive number of patent applications to investigate the \textit{intensive margin}.

\begin{figure}[htbp]
    \centering
    \includegraphics[width=0.95\textwidth]{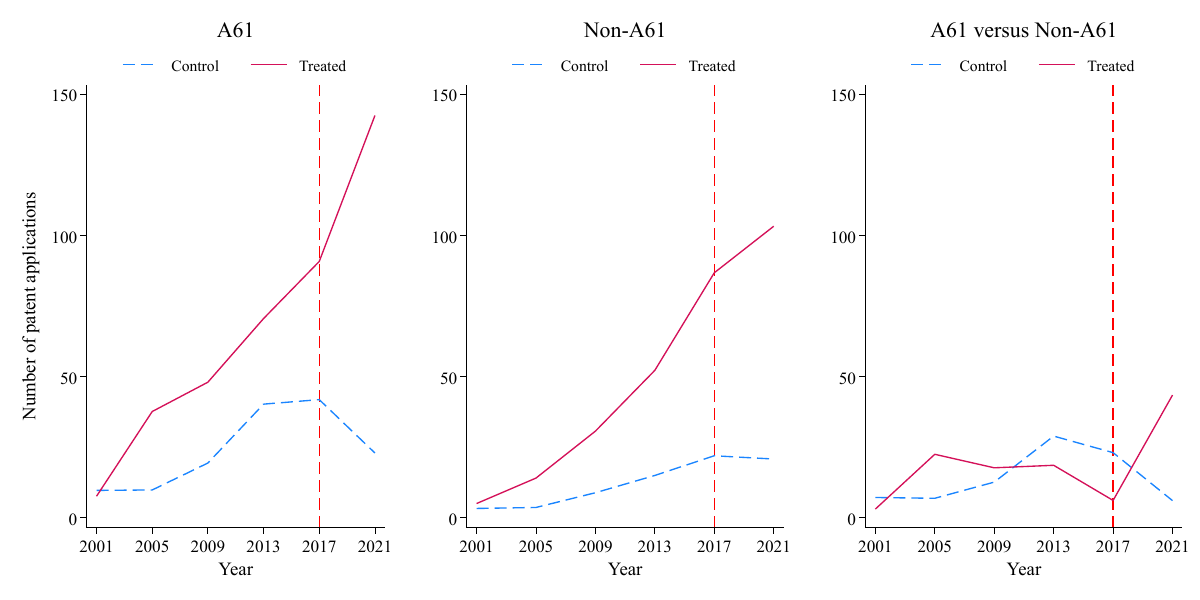}
    \vspace{2mm}
    \caption{When pre-trends also differ across subgroups}
    \label{fig:1}
    \vspace{-2mm}
\begin{fignote}
\textit{Notes}: The figure shows how the numbers of patent applications in A61 and non-A61 tracks evolve over time.
    \vspace{-2mm}
\end{fignote}
\end{figure}

However, as shown by \autoref{fig:1}, neither A61 nor non-A61 tracks exhibit parallel trends in the outcome variable due to its exponential growth in value in the treated ``4+7'' cities, and the parallel trends are likely violated not only along the treatment dimension but also along the dimension of patent track. While synthetic DID can help create a more accurate counterfactual when parallel trends are violated along the treatment dimension, it might not capture complexities across different subgroups in this situation. As shown in \autoref{fig:a1}, pre-treatment trends are not as parallel as we may want based on synthetic DID. Thus, my derivations above open an avenue for synthetic DDD, which ensures that the counterfactuals for both the treatment and the control groups are robust and account for these multi-dimensional violations of parallel trends. There can also be complex interactions between treatment, time, and subgroups that cannot be fully captured by synthetic DID alone. Thus, situations like these motivate us to combine SC with DDD.\footnote{A caution about applying the synthetic DDD procedure to count data is given in Section \ref{applicability}.}

\subsection{How?}

We can write out the empirical equations for regular DDD and transformed DDD first:
\begin{align}
    \begin{split}
        \text{NumPatents}_{it} = & \gamma_0 + \gamma_1 (\text{Treat}_i \times \text{Post}_t \times \text{Med}_i) + \gamma_2 (\text{Treat}_i \times \text{Post}_t) +\\
        & \gamma_3 (\text{Post}_t \times \text{Med}_i) + X^{\prime}_{it} \gamma_4 + \mu_i + \lambda_t + \epsilon_{it},
    \end{split}
\end{align}
\begin{align}
    \begin{split}
        \text{DiffNumPatents}_{it} = & \theta_0 + \theta_1 (\text{Treat}_i \times \text{Post}_t) + \mu_i + \lambda_t + e_{it}, \label{eq:empiricaldiff}
    \end{split}
\end{align}
where $\text{Med}_i$ takes the value of 1 if a track is related to A61 and 0 otherwise, and $\text{DiffNumPatents}_{it}$ is the difference in the number of A61 patent applications and the predicted average number of non-A61 patent applications in a specific year. To see how the transformed DDD performs, we compare results of standard and transformed DDD estimations without and with covariates in \autoref{tab:compare}.

\begin{table}[htbp]
\begin{threeparttable}
\caption{Performance of regular and transformed DDD estimations}
\label{tab:compare}
\footnotesize
\begin{tabularx}{\textwidth}{@{}l*{4}{>{\centering\arraybackslash}X}@{}}
\toprule
& \multicolumn{2}{c}{Without covariates} & \multicolumn{2}{c}{With covariates} \\
\cmidrule(lr){2-3} \cmidrule(lr){4-5}
 & (1) & (2) & (3) & (4) \\
 & Standard & Transformed & Standard & Transformed \\
\midrule
Treatment effect & 37.586 & 37.586 & 39.336 & 39.567 \\
Clustered robust SE & 20.942 & 19.287 & 20.632 & 19.117 \\
Robust P-value & 0.073 & 0.052 & 0.057 & 0.039 \\
Regular SE & 14.614 & 8.949 & 14.572 & 8.883 \\
Regular P-value & 0.010 & 0.000 & 0.007 & 0.000 \\
\bottomrule
\end{tabularx}
\begin{tablenotes}
\item \scriptsize \textit{Notes}: The table shows the effects of ``4+7 Cities'' Centralized Drug Procurement pilot program on the annual number of patent applications at the track level based on 2,058 observations; SE stands for standard errors, and the clustered ones are clustered at the track level; *** p$<$0.01, ** p$<$0.05, * p$<$0.1; covariates include city-level population, GDP, fiscal revenue, fiscal expenditure, R\&D spending, educational spending, number of large-scale industrial firms, number of hospitals, number of medical beds, and number of physicians.
\end{tablenotes}
\end{threeparttable}
\end{table}

As shown in the table, when we do not include any covariates, the transformed DDD estimation gives exactly the same point estimate of the treatment effect as the standard DDD does, although standard errors are different. They differ due to several factors related to how each model handles residual variance and degrees of freedom. For instance, regular DDD involves a larger number of interaction terms, such as $\text{Treat}_i \times \text{Post}_t \times G_i$, which increases the complexity of the residual variance. This complexity can lead to a different calculation of the error variance, particularly if there is variation in the subgroup interaction terms. The transformed DDD simplifies the model by transforming the outcome into first differences. This transformation can alter the distribution of the residuals, potentially leading to a more efficient estimator with lower standard errors. The simplification reduces the number of parameters to estimate, as fewer interaction terms are involved, which can affect how the standard errors are computed.

Since DID tends to over-reject the null hypothesis by having a smaller standard error \citep{bertrand2004}, one may use clustered robust standard errors to address this issue. In Stata, the regular DDD can be estimated using command ``reghdfe'', while the transformed DDD can be estimated using command ``xtreg'' or ``reghdfe''.

When including covariates, however, point estimates can differ. This can be due to finite sample sizes, especially if the number of observations in certain groups is small. Nevertheless, my empirical exercise shows that strictly following Equations (\ref{eq:fit}) to (\ref{eq:w2}) yields the closest point estimates.

\begin{figure}[ht]
    \centering
    \includegraphics[width=0.95\textwidth]{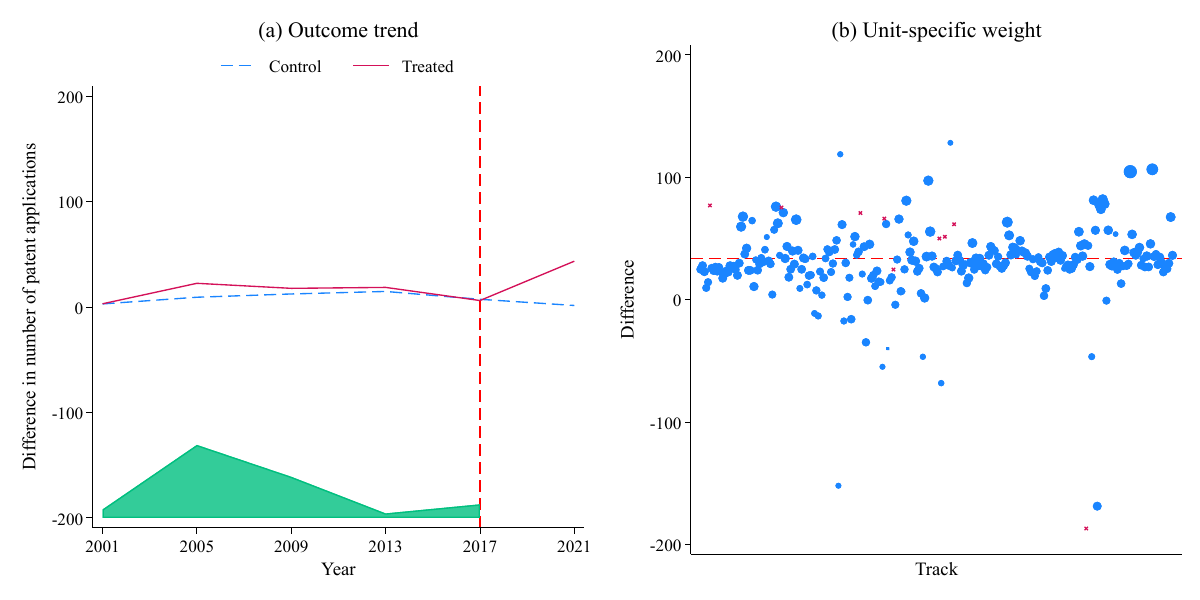}
    \vspace{2mm}
    \caption{When combining SC with transformed DDD}
    \label{fig:2}
    \vspace{-2mm}
\begin{fignote}
\textit{Notes}: The figure shows how the difference in the numbers of A61 patent applications and the predicted average number of non-A61 patent applications evolves over time and is affected by the Centralized Drug Procurement pilot program. The size of the blue dots on the right panel reflect the weights assigned to a specific track, and the little red cross indicates that the weight is zero.
    \vspace{-2mm}
\end{fignote}
\end{figure}

Then, we can follow \citet{arkhangelsky2021} to combine the transformed DDD with SC using the Stata command ``sdid''. From \autoref{fig:2}, we can tell that parallel trends are much more apparent than those in \autoref{fig:1} after the SC weighting. The point etimate suggests a treatment effect of 33.6 more patent applications in A61 track relative to non-A61 track annually. The standard error based on \citet{arkhangelsky2021}'s placebo procedure is only about 5.9, leading to a P-value smaller than 0.001. We may conclude that centralized drug procurement can promote innovation activities in terms of the number of patent applications.

\section{Additional Notes}

This section presents a more formal description of the synthetic DDD estimator. I describe the estimator, the underlying assumptions, and derive its large sample distribution, emphasizing the use of the transformed outcome variable $W_{itjx}$.

\subsection{The synthetic DDD estimator}

I extend the synthetic DID framework to accommodate the triple-difference setting, capturing heterogeneous treatment effects across subgroups. The estimator relies on constructing a counterfactual outcome for the treated units that uses a weighted combination of control units and pre-treatment periods. The transformation we apply to the outcome is described in Equations (\ref{eq:fit}) and (\ref{eq:w2}). Let $\omega_i$ denote the unit-specific weights for the control group and $\lambda_t$ represent the time-specific weights for pre-treatment periods. The synthetic DDD (SDDD) estimator is then defined as:
\begin{equation}
(\hat{\delta}^{\text{SDDD}}, \hat{\alpha}, \hat{\mu}, \hat{\theta}) = \arg\min_{\delta, \alpha, \mu, \theta} \left\{ \sum_{i \in \{i \mid G_i = 1\}, t} \left(W_{itjx} - \alpha - \mu_i - \theta_t - D_{it} \delta \right)^2 \hat{\omega}_i^{sdid} \hat{\lambda}_t^{sdid} \right\}, \label{eq:SDDD}
\end{equation}
where $\hat{\omega}_i^{sdid}$ and $\hat{\lambda}_t^{sdid}$ are chosen based on \citet{arkhangelsky2021}'s procedure.

\subsection{Key assumptions}

To ensure that the proposed estimator provides consistent and unbiased estimates of the treatment effect, two important assumptions should be noted:

\begin{enumerate}
    \item \textbf{Large Panel Assumption}: The estimator assumes a large panel, where both the number of cross-sectional units \( N \) and the number of time periods \( T \) grow large. This is necessary for consistency and asymptotic normality.
    \item \textbf{Normality of Error Terms}: The error terms \( \varepsilon_{it} \) are assumed to be normally distributed. This is important for deriving the large sample distribution.
\end{enumerate}

\subsection{Large sample distribution}

Following the methodology of \cite{arkhangelsky2021}, one can derive the large sample distribution of the synthetic DDD estimator. Under the assumption that both \( N \to \infty \) and \( T \to \infty \), the synthetic DDD estimator converges to a normal distribution. Specifically, we have:
\begin{equation}
\sqrt{NT} \left( \hat{\delta}_{\text{SDDD}} - \delta \right) \xrightarrow{d} \mathcal{N}(0, \sigma^2),
\end{equation}
where \( \delta \) is the true treatment effect and \( \sigma^2 \) is the asymptotic variance of the estimator. 

The variance \( \sigma^2 \) is determined by the residual variance and the structure of the weights. The SC approach minimizes the differences between treated and control units in the pre-treatment period using weights.

\subsection{Applicability to the empirical setting} \label{applicability}

In the empirical application above, I have only six time periods, which challenges the large sample assumption. In practice, finite sample sizes can affect the precision of the estimates. To account for this, one may use a block bootstrap procedure to estimate the variance in cases where the number of time periods is limited. 

Another challenge the empirical exercise faces is the assumption that the error terms are normally distributed. However, one may argue that even though the number of patent applications is count data, when aggregated over time or across regions, the Central Limit Theorem implies that the distribution of these aggregated counts may approximate normality, making the assumption of normally distributed error terms reasonable in large samples.

Nevertheless, the empirical example provided earlier may only serve as a working illustration of when and how to apply the synthetic DDD strategy. Researchers should always be cautious about the underlying assumptions for an appropriate use of SC in a DDD setting.

\subsection{Why don't we combine SC with DDD directly?}

One might ask: why don't we create a SDDD estimator by directly combining SC with triple difference in the following form:\footnote{Here I do not include covariates since we can simply incorporate adjustment for them by applying SDDD to the residuals $Y_{it}^{Res} = Y_{it} - X_{it}^{\prime}\hat{\beta}$ of the regression of $Y_{it}$ on $X_{it}$.}
\begin{equation}
    (\hat{\delta}^{\text{sddd}}, \hat{\alpha}, \hat{\gamma}_1, \hat{\gamma}_2, \hat{\mu}, \hat{\theta}) = \arg\min_{\delta, \alpha, \gamma_1, \gamma_2 \mu, \theta} \left\{ \sum_{i, t}  \left(\begin{aligned} 
    & Y_{it} - \alpha \\
    & - \delta (\text{Treat}_i \times \text{Post}_t \times G_i) \\
    & - \gamma_1 (\text{Treat}_i \times \text{Post}_t) \\
    & - \gamma_2 (\text{Post}_t \times G_i) \\
    & - \mu_i - \theta_t\end{aligned}\right)^2 \hat{\omega}_i^{sddd} \hat{\lambda}_t^{sddd} \right\} \label{eq:directSDDD}
\end{equation}
where  $\hat{\omega}_i^{sddd}$ and $\hat{\lambda}_t^{sddd}$ should be chosen based on a new procedure.

At first glance, this approach seems intuitive. The SC method could be used to construct weights $\hat{\omega}_i^{sddd}$ and $\hat{\lambda}_t^{sddd}$ for both the cross-sectional units and the time periods, helping to create a more accurate counterfactual for treated units. The interaction terms $(\text{Treat}_i \times \text{Post}_t \times G_i)$, $(\text{Treat}_i \times \text{Post}_t)$, and $(\text{Post}_t \times G_i)$ could account for the heterogeneous treatment effects across subgroups in the DDD setup. However, there are several challenges and considerations that may prevent this approach from being ideal, or even necessary, in practice:
\begin{enumerate}
    \item \textbf{Computational Complexity}: Equation (\ref{eq:directSDDD}) involves much more complex computations compared to Equation (\ref{eq:SDDD}). Estimating the weights $\hat{\omega}_i^{sddd}$ for each cross-sectional unit, $\hat{\lambda}_t^{sddd}$ for each time period, and the parameters $\hat{\delta}^{\text{sddd}}$, $\hat{\alpha}$, $\hat{\gamma}_1$, $\hat{\gamma}_2$, $\hat{\mu}$, and $\hat{\theta}$ simultaneously would involve a much larger optimization problem. The increased dimensionality could lead to issues with overfitting, unstable weights, and increased sensitivity to the choice of covariates or model specification.
    \item \textbf{Interpretability}: The interpretation of the weights becomes more complex when combining SC directly with DDD. In Equation (\ref{eq:SDDD}), the weights represent how much each control unit or time period contributes to constructing the counterfactual for treated units in a DID setting. However, in Equation (\ref{eq:directSDDD}), treatment effects are heterogeneous across both subgroups and time periods. Constructing appropriate weights that reflect these multi-dimensional interactions becomes challenging. It may be difficult to determine whether the weights provide meaningful interpretations for policy analysis and whether they accurately capture the multi-dimensional heterogeneity that DDD models aim to address.
    \item \textbf{Potential Redundancy}: The transformed DDD approach discussed earlier already simplifies the DDD estimator by reducing it to a DID structure. By transforming the outcome variable to account for differences between treated and control subgroups, we can apply the synthetic DID framework to the transformed outcome without explicitly constructing a new synthetic DDD estimator. This transformed approach effectively addresses violations of parallel trends across both time and subgroup dimensions, as SC constructs a more accurate counterfactual based on the adjusted outcome. Thus, directly combining SC with DDD might be redundant when the transformed DDD approach already achieves the same goal.
\end{enumerate}

\section{Concluding Remarks}

This paper introduces a practical way to address the parallel trends assumption in DDD models by transforming them into a DID framework and then combining them with SC, in order to improve counterfactual estimation when parallel trends are likely violated across multiple dimensions. This approach is especially relevant in evaluating complex policies such as China’s ``4+7 Cities'' Centralized Drug Procurement pilot program, where treatment effects may differ significantly between subgroups.

The introduction of SC in DDD settings addresses a key gap in the literature. While balancing methods have been widely applied in DID contexts to account for violations of parallel trends, their application to DDD has been largely unexplored. For empirical researchers, this method expands the use of SC to capture subgroup-specific variations in treatment effects, which are often missed by traditional DDD estimators, and allows for more precise estimation of treatment effects in complex empirical settings where interactions between subgroups and treatment are likely to influence the outcome. By offering a more flexible and robust framework, this approach encourages researchers to move beyond the standard DDD assumptions when subgroup-specific violations of parallel trends are suspected.

From a policy evaluation perspective, the application of synthetic DDD to the ``4+7'' program highlights a more credible consequence on high-quality development. The program, designed to reduce drug prices and promote the use of generics, affects different cities and industries in varying ways. Standard DDD may overstate its effect on innovation, particularly when pre-trends differ across subgroups. Synthetic DDD, by creating more accurate counterfactuals, offers a clearer understanding of how the program has impacted pharmaceutical innovation and competition. As shown by the figures, the effect becomes more modest if we account for the differential pre-trends.

While this paper demonstrates the efficacy of combining SC with DDD, there are opportunities for future research. Further work could explore its application in other policy contexts or investigate alternative weighting methods to enhance the performance of SC in DDD settings with complex data structures. This combination of SC and DDD offers a powerful tool for empirical research, expanding the scope of more reliable policy evaluation.

\bibliography{reference}

\newpage
\begin{appendices}
\setcounter{table}{0}
\setcounter{figure}{0}
\setcounter{equation}{0}
\renewcommand{\thetable}{A\arabic{table}}
\renewcommand{\thefigure}{A\arabic{figure}}
\renewcommand{\theequation}{A.\arabic{equation}}

\section{Proofs} \label{proofs}
\subsection{Transforming the ATT in a simple DID} \label{proof:att}

The ATT for a simple DID as expressed in Equation (\ref{eq:att}) can be written as
\begin{align}
\begin{split}
    \alpha_1 = & \left( \mathbb{E}[Y_{it} \mid \text{Treat}_i = 1, \text{Post}_t = 1] - \mathbb{E}[Y_{it} \mid \text{Treat}_i = 1, \text{Post}_t = 0] \right) - \\
    & \left( \mathbb{E}[Y_{it} \mid \text{Treat}_i = 0, \text{Post}_t = 1] - \mathbb{E}[Y_{it} \mid \text{Treat}_i = 0, \text{Post}_t = 0] \right) \\
    = & \left( \mathbb{E}[Y_{it} \mid \text{Treat}_i = 1, \text{Post}_t = 1] - \mathbb{E}[Y_{it} \mid \text{Treat}_i = 0, \text{Post}_t = 1] \right) - \\
    & \left(\mathbb{E}[Y_{it} \mid \text{Treat}_i = 1, \text{Post}_t = 0] - \mathbb{E}[Y_{it} \mid \text{Treat}_i = 0, \text{Post}_t = 0] \right) \\
    = & \mathbb{E}\left[Y_{it}  - \mathbb{E}[Y_{it} \mid \text{Treat}_i = 0, \text{Post}_t = 1] \mid \text{Treat}_i = 1, \text{Post}_t = 1\right] - \\
    & \mathbb{E}\left[Y_{it} - \mathbb{E}[Y_{it} \mid \text{Treat}_i = 0, \text{Post}_t = 0] \mid \text{Treat}_i = 1, \text{Post}_t = 0\right] \\
    = & \mathbb{E}\left[ \mathbb{E}\left[Y_{it}  - \mathbb{E}[Y_{it} \mid \text{Treat}_i = 0, t = t] \mid \text{Post}_t = 1\right] - \right.\\
    & \left.\mathbb{E}\left[Y_{it} - \mathbb{E}[Y_{it} \mid \text{Treat}_i = 0, t = t] \mid \text{Post}_t = 0\right] \mid \text{Treat}_i = 1 \right] \\
    = & \mathbb{E}\left[ \mathbb{E}\left[Z_{it} \mid \text{Post}_t = 1\right] - \mathbb{E}\left[Z_{it} \mid \text{Post}_t = 0\right] \mid \text{Treat}_i = 1 \right] \\
    = &  \mathbb{E}\left[Z_{it} \mid \text{Post}_t = 1\right] - \mathbb{E}\left[Z_{it} \mid \text{Post}_t = 0\right] \text{ if } \text{Treat}_i = 1
\end{split}
\end{align}
where $Z_{it}$ is defined by Equation (\ref{eq:z}).

\newpage
\subsection{Transforming the triple difference effect} \label{proof:ddd}

The DDD effect is given by:
\begin{align}
    \begin{split}
        \delta = &\left[ (\mathbb{E}[Y_{it} \mid \text{T}_i = 1, \text{P}_t = 1, \text{G}_i = 1] - \mathbb{E}[Y_{it} \mid \text{T}_i = 1, \text{P}_t = 0, \text{G}_i = 1]) \right. \\
        &- \left. (\mathbb{E}[Y_{it} \mid \text{T}_i = 0, \text{P}_t = 1, \text{G}_i = 1] - \mathbb{E}[Y_{it} \mid \text{T}_i = 0, \text{P}_t = 0, \text{G}_i = 1]) \right] \\
        &- \left[ (\mathbb{E}[Y_{it} \mid \text{T}_i = 1, \text{P}_t = 1, \text{G}_i = 0] - \mathbb{E}[Y_{it} \mid \text{T}_i = 1, \text{P}_t = 0, \text{G}_i = 0]) \right. \\
        &- \left. (\mathbb{E}[Y_{it} \mid \text{T}_i = 0, \text{P}_t = 1, \text{G}_i = 0] - \mathbb{E}[Y_{it} \mid \text{T}_i = 0, \text{P}_t = 0, \text{G}_i = 0]) \right] \\
        = &\left[ (\mathbb{E}[Y_{it} \mid \text{T}_i = 1, \text{P}_t = 1, \text{G}_i = 1] - \mathbb{E}[Y_{it} \mid \text{T}_i = 1, \text{P}_t = 1, \text{G}_i = 0]) \right. \\
        &- \left. (\mathbb{E}[Y_{it} \mid \text{T}_i = 1, \text{P}_t = 0, \text{G}_i = 1] - \mathbb{E}[Y_{it} \mid \text{T}_i = 1, \text{P}_t = 0, \text{G}_i = 0]) \right] \\
        &- \left[ (\mathbb{E}[Y_{it} \mid \text{T}_i = 0, \text{P}_t = 1, \text{G}_i = 1] - \mathbb{E}[Y_{it} \mid \text{T}_i = 0, \text{P}_t = 1, \text{G}_i = 0]) \right. \\
        &- \left. (\mathbb{E}[Y_{it} \mid \text{T}_i = 0, \text{P}_t = 0, \text{G}_i = 1] - \mathbb{E}[Y_{it} \mid \text{T}_i = 0, \text{P}_t = 0, \text{G}_i = 0]) \right] \\
        = &\left[ \mathbb{E}\left[Y_{it} - \mathbb{E}[Y_{it} \mid \text{T}_i = j, t = t, \text{G}_i = 0] \mid \text{T}_i = 1, \text{P}_i = 1, \text{G}_i = 1\right] \right. \\
        &- \left. \mathbb{E}\left[Y_{it} - \mathbb{E}[Y_{it} \mid \text{T}_i = j, t = t, \text{G}_i = 0] \mid \text{T}_i = 1, \text{P}_i = 0, \text{G}_i = 1 \right] \right] \\
        &- \left[ \mathbb{E}\left[Y_{it} - \mathbb{E}[Y_{it} \mid \text{T}_i = j, t = t, \text{G}_i = 0] \mid \text{T}_i = 0, \text{P}_i = 1, \text{G}_i = 1 \right] \right. \\
        &- \left. \mathbb{E}\left[Y_{it} - \mathbb{E}[Y_{it} \mid \text{T}_i = j, t = t, \text{G}_i = 0] \mid \text{T}_i = 0, \text{P}_i = 0, \text{G}_i = 1 \right] \right] \\
        = &\mathbb{E}\left[ \left[ \mathbb{E}\left[Y_{it} - \mathbb{E}[Y_{it} \mid \text{T}_i = j, t = t, \text{G}_i = 0] \mid \text{T}_i = 1, \text{P}_i = 1 \right] \right. \right. \\
        &- \left. \mathbb{E}\left[Y_{it} - \mathbb{E}[Y_{it} \mid \text{T}_i = j, t = t, \text{G}_i = 0] \mid \text{T}_i = 1, \text{P}_i = 0\right] \right] \\
        &- \left[ \mathbb{E}\left[Y_{it} - \mathbb{E}[Y_{it} \mid \text{T}_i = j, t = t, \text{G}_i = 0] \mid \text{T}_i = 0, \text{P}_i = 1 \right] \right. \\
        &- \left. \left. \mathbb{E}\left[Y_{it} - \mathbb{E}[Y_{it} \mid \text{T}_i = j, t = t, \text{G}_i = 0] \mid \text{T}_i = 0, \text{P}_i = 0 \right] \right] \mid \text{G}_i = 1 \right] \\
        = &\mathbb{E}\left[ \left[ \mathbb{E}\left[W_{itj} \mid \text{T}_i = 1, \text{P}_i = 1 \right] - \mathbb{E}\left[W_{itj} \mid \text{T}_i = 1, \text{P}_i = 0 \right] \right] \right. \\
        &- \left. \left[ \mathbb{E}\left[W_{itj} \mid \text{T}_i = 0, \text{P}_i = 1 \right] - \mathbb{E}\left[W_{itj} \mid \text{T}_i = 0, \text{P}_i = 0 \right] \right] \mid \text{G}_i = 1 \right] \\
        = & \left[ \mathbb{E}\left[W_{itj} \mid \text{T}_i = 1, \text{P}_i = 1 \right] - \mathbb{E}\left[W_{itj} \mid \text{T}_i = 1, \text{P}_i = 0 \right] \right] \\
        &- \left[ \mathbb{E}\left[W_{itj} \mid \text{T}_i = 0, \text{P}_i = 1 \right] - \mathbb{E}\left[W_{itj} \mid \text{T}_i = 0, \text{P}_i = 0 \right] \right] \text{ if } \text{G}_i = 1 
    \end{split}
\end{align}
where $W_{itj}$ is defined by Equation (\ref{eq:w}), T stands for Treat, and P stands for Post.

Apparently, this is equivalent to $\theta_1$ in the following model for individuals in the subgroup of interest ($\text{G}_i = 1$):
\begin{equation}
    W_{it} = \theta_0 + \theta_1 \text{D}_t + \mu_i + \lambda_t + e_{it} \text{ if } G_i = 1.
\end{equation}
where $D_{it} = \text{Treat}_i \times \text{Post}_t$ is the DID interaction term.

Under Assumptions \ref{ass:parallel2} and \ref{ass:exogeneity2}, $\hat{\theta}_1$ is a consistent estimator of $\delta$:
\begin{align}
    \begin{split}
        \hat{\theta}_1 = & \left[\left( \hat{\bar{Y}}_{\text{Treat}, \text{Post}, \text{Target}} - \hat{\bar{Y}}_{\text{Treat}, \text{Post}, \text{Non-target}} \right) \right. \\
        - & \left. \left( \hat{\bar{Y}}_{\text{Treat}, \text{Pre}, \text{Target}} - \hat{\bar{Y}}_{\text{Treat}, \text{Pre}, \text{Non-target}} \right) \right] \\
        - & \left[\left( \hat{\bar{Y}}_{\text{Control}, \text{Post}, \text{Target}} - \hat{\bar{Y}}_{\text{Control}, \text{Post}, \text{Non-target}} \right) \right. \\
        - & \left. \left( \hat{\bar{Y}}_{\text{Control}, \text{Pre}, \text{Target}} - \hat{\bar{Y}}_{\text{Control}, \text{Pre}, \text{Non-target}} \right) \right] \xrightarrow{p} \alpha_1
    \end{split}
\end{align}
where $\hat{\bar{Y}}_{\text{Group}, \text{Time}, \text{Subgroup}}$ denotes sample averages for Group (either treatment or control groups), Time (either pre-treatment or post-treatment years), and Subgroup (either target or non-target subgroups), and ``$\xrightarrow{p}$'' denotes convergence in probability.

\newpage
\section{More Details about the Policy} \label{appendix:policy}

The ``4+7 Cities'' Centralized Drug Procurement pilot program was initiated in late 2018 as part of a broader effort to reform the pharmaceutical sector and lower drug prices in China. The pilot aimed to centralize the procurement of drugs through competitive bidding, allowing the government to negotiate lower prices directly with pharmaceutical companies. This initiative initially targeted 11 cities, hence the ``4+7''---i.e., four municipalities (Beijing, Shanghai, Chongqing, and Tianjin) and seven major cities (Guangzhou, Shenzhen, Xi'an, Dalian, Chengdu, Shenyang, and Xiamen).

The program's goal was twofold:
\begin{itemize}
    \item \textit{Reduce drug prices}: By creating a centralized bidding system, the government could leverage its bargaining power to obtain lower prices for selected drugs, which were often essential medications.
    \item \textit{Promote the use of generic drugs}: The program favored the procurement of high-quality generic drugs over brand-name alternatives, encouraging pharmaceutical companies to invest in generics and promoting competition in the drug market.
\end{itemize}
Under the ``4+7'' program, companies that won the bidding process were granted the exclusive right to supply certain drugs to public hospitals in the pilot cities. In return, they had to commit to offering substantial price reductions—sometimes by more than 50\%. The pilot's success led to its expansion to additional regions in China, and it has become a key component of the government's broader health care reform initiatives.

In terms of impact, the program has been found to successfully reduce drug prices and improve access to essential medicines. However, concerns have also been raised about potential effects on pharmaceutical innovation and the profitability of smaller drug manufacturers, particularly those who lose out in the bidding process.

\newpage
\section{More Details about Data} \label{appendix:data}

I manually collect patent-level records from China National Intellectual Property Administration. In each record, I have the information of location (specific to a city), submission date, release date, IPC symbol, inventor type (i.e., firm, school, individual, and government and others), and patent type (i.e., invention versus utility model). Then, I aggregate the data by counting the number of patent applications in each submission year for each track, which is defined as a ``city, four-digit IPC code, patent type, and inventor type'' pair. 

Since the records of most patent applications are released within three years, by the time (June 2024) I compiled the data, the most recent year with almost complete public records was 2021. Thus, I only include year 2021 as a post-treatment year. Then, I collect data every four year, leading to a panel data with equal space between years, spanning from 2001 to 2021. One key reason for skipping 2019-2020 is to avoid the possible impact of COVID-19 and my limited access to 2019-2020 data. In 2021, the numbers of COVID-19 cases were low in China, and there were basically no lockdowns or quarantine within the country. Moreover, since the pilot program was initiated in late 2018, I create a ``donut'' estimation by excluding the year. This leads to six waves of data in 2001, 2005, 2009, 2013, 2017, and 2021.

According to World Intellectual Property Organization, patents whose IPC symbol starts with ``A61'' are related to medical or veterinary science. For example, A61B patents are related to diagnosis, surgery, and identification; A61C patents are related to dentistry, apparatus or methods for oral or dental hygiene; A61D patents are about veterinary instruments, implements, tools, or methods; for more details about A61F, A61G, A61H, A61J, A61K, A61L, A61M, A61N, A61P, and A61Q, please rever to \url{https://www.wipo.int/classifications/en/}.

\newpage
\section{An Additional Result}

\begin{figure}[ht]
    \centering
    \includegraphics[width=0.95\textwidth]{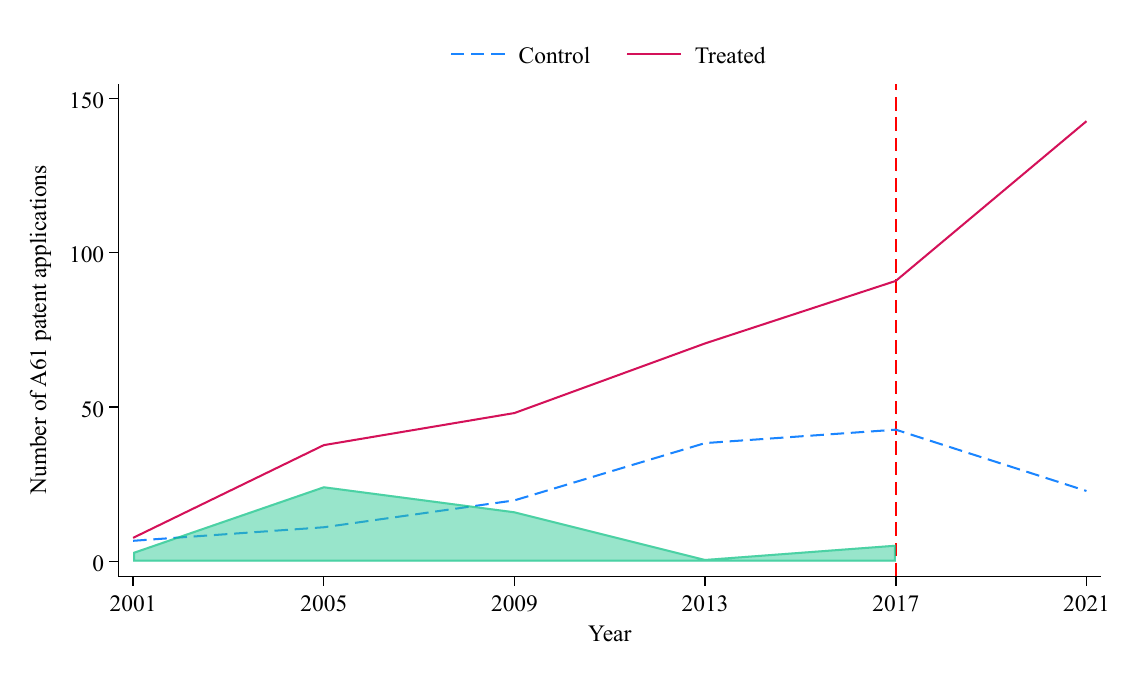}
    \vspace{2mm}
    \caption{When the parallel trends of synthetic DID are not good enough}
    \label{fig:a1}
    \vspace{-2mm}
\begin{fignote}
\textit{Notes}: The figure shows how the number of A61 patent applications evolves over time and is affected by the Centralized Drug Procurement pilot program based on a synthetic DID regression with covariates listed in \autoref{tab:compare} using A61 tracks. As shown, the pre-treatment trends are not parallel even after synthetic control.
    \vspace{-2mm}
\end{fignote}
\end{figure}

\end{appendices}

\end{document}